\documentclass{pasj00}

\begin{document}
\SetRunningHead{Author(s) in page-head}{Running Head}
\Received{2003/07/??}
\Accepted{??}

\title{Magnetically arrested disk: an energetically efficient
accretion flow}

\author{Ramesh \textsc{Narayan}}
\affil{Harvard-Smithsonian Center for Astrophysics, 60 Garden Street,
Cambridge, MA 02138, USA}
\email{rnarayan@cfa.harvard.edu}

\author{Igor V. \textsc{Igumenshchev}}
\affil{Laboratory for Laser Energetics, University of Rochester,
250 East River Road, Rochester, NY 14623, USA} 
\email{iigu@lle.rochester.edu}
\and
\author{Marek A. \textsc{Abramowicz}}
\affil{Department of Astronomy and Astrophysics, G\"oteborg
University and} 
\affil{Chalmers University of Technology, S-41296, G\"oteborg, Sweden}
\email{marek@fy.chalmers.se}

%


\maketitle

\begin{abstract}

We consider an accretion flow model originally proposed by
Bisnovatyi-Kogan \& Ruzmaikin (1974), which has been confirmed in
recent 3D MHD simulations.  In the model, the accreting gas drags in a
strong poloidal magnetic field to the center such that the accumulated
field disrupts the axisymmetric accretion flow at a relatively large
radius.  Inside the disruption radius, the gas accretes as discrete
blobs or streams with a velocity much less than the free-fall
velocity.  Almost the entire rest mass energy of the gas is released
as heat, radiation and mechanical/magnetic energy.  Even for a
non-rotating black hole, the efficiency of converting mass to energy
is of order 50\% or higher.  The model is thus a practical analog of
an idealized engine proposed by Geroch and Bekenstein.

\end{abstract}

\section{Introduction}

Following a suggestion by Geroch (colloquium, Princeton Univ.,
Dec. 1971), Bekenstein (1972) described an engine that makes use of
the extreme gravitational potential of a black hole (BH) to convert
mass to energy with nearly perfect efficiency.  The engine works by
slowly lowering a mass $m$ into the BH potential using a strong wire.
As the mass is lowered, the energy $E$ as measured at infinity
decreases relative to the initial energy $E_0=mc^2$.  The change in
energy is equal to the mechanical work done by the wire back at the
engine.  For a non-rotating BH, if the mass is lowered from an
initially large radius down to a final radius $R$, and if the mass is
then allowed to fall freely into the BH, the efficiency of the engine
is
\begin{equation}
\eta =\left[E_0-E(r)\right]/mc^2 = \left[ 1- (1-1/r)^{1/2} \right].
\end{equation}
Here $r=R/R_S$ is the radius in Schwarzschild units, where
$R_S=2GM/c^2$ and $M$ is the BH mass.  As $r \to 1$ all the rest mass
energy of $m$ is converted into work, and $\eta \to 1$.

The Geroch-Bekenstein engine is not believed to occur naturally in
astrophysical systems.  Astrophysical BHs do convert mass to energy,
but they do it via accretion flows (see Kato, Fukue \& Mineshige 1998
for a review), which generally have modest efficiencies.  We describe
in \S2 a kind of accretion flow that was originally discussed by
Bisnovatyi-Kogan \& Ruzmaikin (1974, 1976, hereafter BKR74, BKR76) and
that has been seen in recent computer MHD simulations.  We argue in
\S3 that this flow has a high efficiency, and we conclude in \S4 with
a brief discussion.

\section{Magnetically Arrested Disk}

Figure 1a shows the basic idea.  We assume that a significant amount
of poloidal magnetic flux has collected in the vicinity of the BH as a
result of the cumulative action of the accretion flow, and that the
magnetic field is dynamically dominant.  The field is prevented from
escaping by the continued inward pressure of accretion.  At the same
time, the field lines do not fall into the BH because the BH only
``wants'' the plasma but ``does not want'' the field (Punsly 2001,
ch. 8).  The accumulated poloidal field disrupts the accretion flow at
a magnetospheric radius $R_m \equiv r_mR_S$, which lies well outside
the event horizon of the BH.  For $r>r_m$, the flow is essentially
axisymmetric, as in any standard accretion flow.  However, for
$r<r_m$, the flow breaks up into blobs or streams, and the gas has to
fight its way towards the BH by a process of magnetic interchanges and
reconnection.  The velocity of the gas in this region is much less
than the free-fall velocity $v_{\rm ff}$.  We call such a disrupted
accretion flow a ``magnetically arrested disk'' or MAD for short.

\begin{figure}
  \begin{center} 
    \FigureFile(60mm,60mm){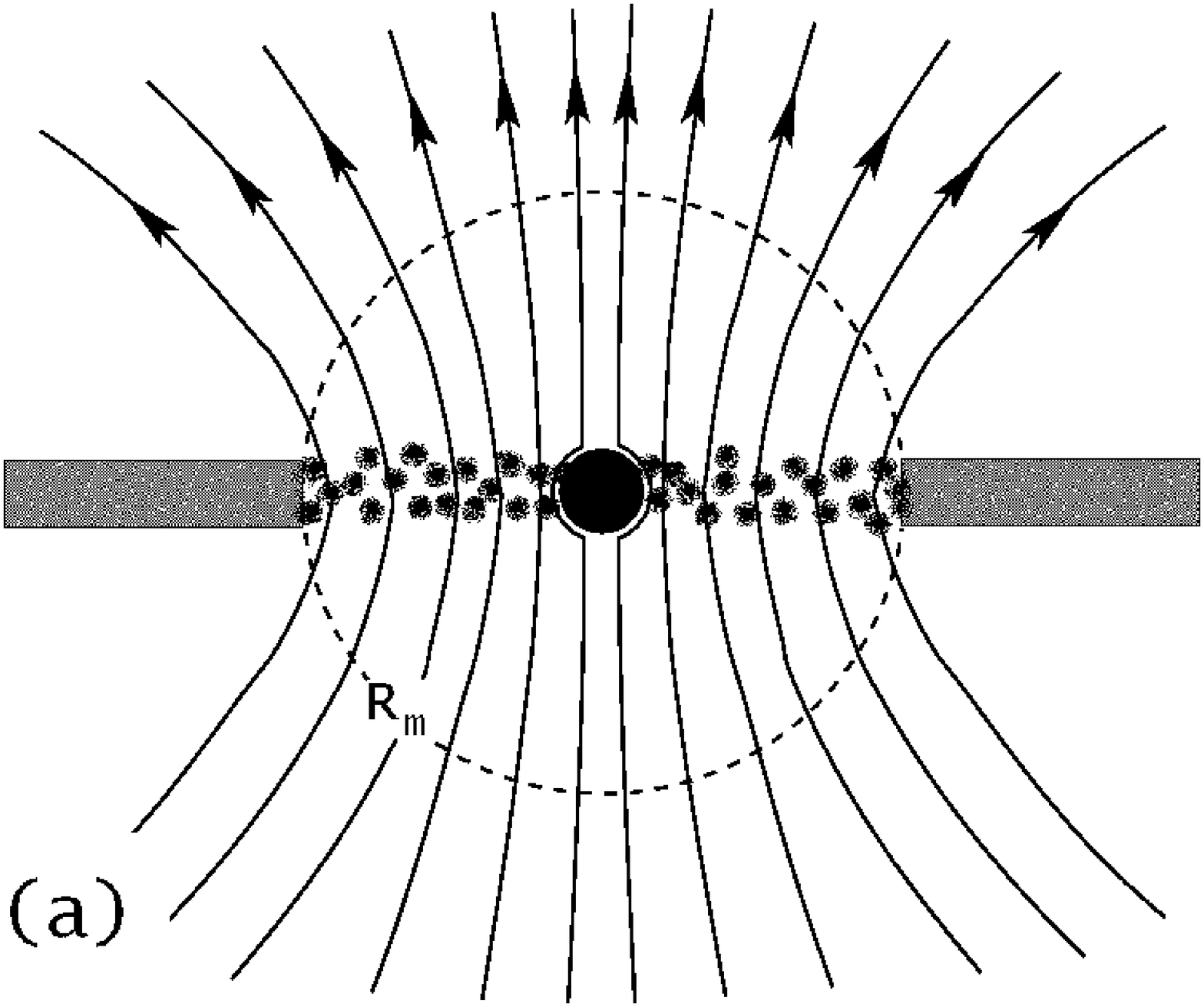} \FigureFile(60mm,60mm){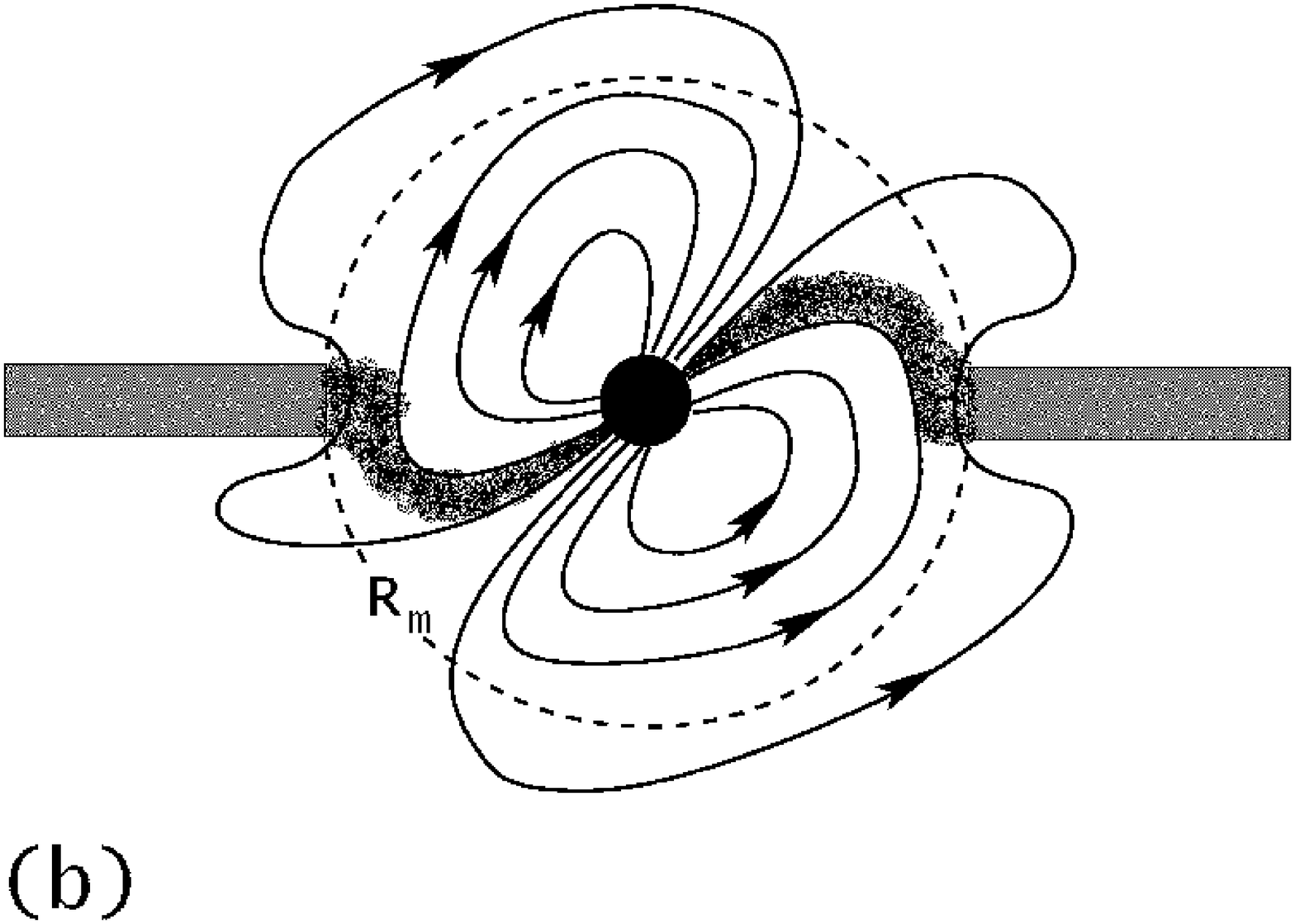}
\end{center} \caption{(a)
Shows the basic elements of the proposed accretion model.  An
axisymmetric accretion disk is disrupted at a magnetospheric radius
$R_m$ by a strong poloidal magnetic field which has accumulated at the
center.  Inside $R_m$ the gas accretes as magnetically confined blobs
which diffuse through the field with a relatively low
velocity. Surrounding the blobs is a hot low-density corona.  (b)
Shows the accretion flow around a magnetized compact star.  An
axisymmetric disk is disrupted at the magnetospheric radius $R_m$ by
the strong stellar field.  Inside $R_m$ the gas follows the magnetic
field lines and free-falls on the polar caps of the star.}\label{fig1}
\end{figure}

A similar idea was discussed by BKR74, who highlighted the importance
of a large-scale dipole magnetic field in a non-rotating accretion
flow (compare their Fig. 1 to our Fig. 1a).  Later, the same authors
constucted a quantitative version of the model (BKR76) using a
prescription for the resistivity of a turbulent plasma analogous to
the $\alpha$-viscosity of turbulent hydrodynamics.  These old ideas
have been confirmed in recent 3D MHD simulations of radiatively
inefficient accretion flows around BHs carried out by us (Igumenshchev
\& Narayan 2002; Igumenshchev, Narayan \& Abramowicz 2003; see
especially Model B in the latter paper).  In the simulations,
frozen-in magnetic flux is dragged to the center by the accreting gas,
causing a substantial increase in the magnetic pressure.  This leads
to a severe disruption of the originally axisymmetric flow, and a
substantial reduction in the gas velocity.

Can we be sure that the magnetic field will become strong enough to
disrupt the accretion flow?  The answer is yes, at least under the
idealized conditions we have considered where the accreting gas brings
in magnetic field of a fixed polarity.  Initially the field near the
BH is not strong enough, so the gas accretes without hindrance.
However, continued accretion keeps bringing in more field and the
magnetic pressure near the BH builds up with time.  Whatever pressure
is needed to disrupt the disk will ultimately be achieved, given
enough time; after that, a MAD flow is inevitable.

The computer simulations mentioned above were done for a radiatively
inefficient accretion flow.  Radiatively efficient disks are sometimes
cool enough to be mostly neutral, in which case the magnetic field
would slip through the gas via ambipolar diffusion, preventing the
accumulation of field at the center.  In addition, there could be
substantial field slippage even in a fully ionized disk, if the
anomalous magnetic diffusivity is large (Lovelace, Romanova \& Newman
1994; Lubow, Papaloizou \& Pringle 1994ab).  However, if magnetic
winds play an important role in the angular momentum loss, then it is
plausible that significant field will be dragged to the center.  In
what follows, we make the optimistic assumption that field-dragging is
efficient and that a configuration as in Fig. 1a does develop.

Magnetically disrupted disks are known in another context, namely
accretion onto magnetized neutron stars and white dwarfs.  A strong
field anchored in the star disrupts the accretion flow at a
magnetospheric radius $R_m \equiv r_m R_S$, as shown in Fig. 1b.  The
topology of the field is, however, very different, since in this case
the gas inside $r\sim r_m$ is able to flow in freely along field lines
down to the magnetic poles of the star.  In the configuration shown in
Fig. 1a, on the other hand, there is no field line connecting gas at
radius $r_m$ to the BH horizon.  The only way for gas to move inward
is by diffusing (via magnetic interchanges) through the strong
magnetic field.  This results in a low velocity and a high energy
efficiency (\S3).

Let us estimate the magnetospheric radius $r_m$ in a MAD flow.  By
assumption, the gas inside $r_m$ moves with a radial component of the
velocity $v_R=\epsilon v_{\rm ff}$, with $\epsilon$ fairly small.  In
the numerical simulations of Igumenshchev et al. (2003), $\epsilon$
was found to be $\sim0.1$.  This value is possibly an overestimate
since the diffusion was dominated by numerical resistivity.  Elsner \&
Lamb (1984), Kaisig, Tajima, \& Lovelace (1992) and Ikhsanov (2001,
and references therein) discuss the physics of diffusion via magnetic
reconnection and interchanges, and show that the diffusion velocity is
given by $v_{\rm diff} \sim \alpha_{\rm R} (\lambda_m/R_m) v_{\rm A}$,
where $\alpha_{\rm R} \sim 0.1 $ is a dimensionless constant, $\lambda
\sim 0.1-0.01 R_m$ is the linear size of reconnection sites, and $v_A$
is the Alfven speed.  Setting $v_{\rm A} \sim v_{\rm ff}$, this gives
$\epsilon = v_{\rm diff} /v_{\rm ff} \sim 10^{-2}-10^{-3}$.  In the
following, we scale all results to $\epsilon_{-2} = \epsilon/10^{-2}$.

The surface density $\Sigma$ of the gas inside $r_m$ is given by
$\Sigma = \dot M/2\pi R\epsilon v_{\rm ff}$, where $\dot M$ is the
mass accretion rate.  Since the magnetic field supports the gas
against gravity, we require $GM\Sigma/R^2 \sim 2B_RB_z/4\pi \sim
B_z^2/2\pi$ (assuming $B_R\sim B_z$).  This gives $B_z \sim
1.5\times10^5 \epsilon_{-2}^{-1/2} m_8^{-1/2}\dot m^{1/2} r^{-5/4}$ G,
where $m_8 \equiv M/10^8M_\odot$, and $\dot m \equiv \dot M/ \dot
M_{\rm Edd}$, with $\dot M_{\rm Edd} = 1.4\times10^{25}m_8 ~{\rm
g\,s^{-1}}$.  Assuming $\epsilon$ is independent of $r$, we may
integrate $B_z$ over radius to estimate the magnetic flux $\Phi$
enclosed within $r_m$.  Inverting this relation, we find
\begin{equation}
r_m \sim 8.2\times10^3 \epsilon_{-2}^{2/3}m_8^{-2}\dot m^{-2/3}
(\Phi/0.1~{\rm pc^2G})^{4/3}.
\end{equation}
The above estimate should be valid for both spherical and rotating
flows, except that $\epsilon$ is perhaps somewhat smaller for rotating
flows.

How much magnetic flux do we expect to collect at the center?  For a
MAD configuration to form it is necessary that the inflowing gas
should have the same sign of $B_z$ (where $z$ is taken to be parallel
to the rotation axis of the flow) for an extended period of time
(BKR74).  In the local ISM near the sun, the magnetic field strength
is a few $\mu$G and the coherence scale of the field is a few hundred
pc or greater.  The magnetic flux in a coherent magnetic patch is thus
$\Phi \sim 0.1 ~{\rm pc^2G}$.  The flux in a spatially coherent patch
in the nucleus of a galaxy is uncertain, but is probably at least
comparable (e.g., Lang, Morris \& Echevarria 1999).  If a flux of this
magnitude is dragged by accretion to the vicinity of a supermassive
BH, equation (2) shows that the accretion disk will be disrupted at
quite a large radius.  Even if only a tiny fraction of the flux is
dragged in, a MAD configuration should still develop.  With continued
accretion, the magnitude of the flux will random-walk, as the field in
successive coherence volumes will be uncorrelated, but the overall
picture should not change.  In a differentially-rotating accretion
disk, additional magnetic field will be generated via the
magneto-rotational instability.  This may lead to stochastic MAD-like
behavior (e.g., Livio, Pringle \& King 2003), with the field reversing
on short time scales.

For an accretion disk in a binary system, the relevant quantities are
the strength of the field in the outer layers of the donor star and
the coherence time of the field in the star, i.e., how long the star
retains a given sign of $B_z$.  This topic is beyond the scope of the
paper.

\section{Efficiency}

We define the efficiency $\eta$ to be the ratio of the energy that
flows out to infinity to the rest energy of the accreting gas
(eq.[1]).  Unlike a standard thin accretion disk, where the
energetically efficient zone lies outside the innermost stable
circular orbit (ISCO) and the gas free-falls inside the ISCO, the gas
in a MAD system has no free-fall region.  The arresting action of the
magnetic field ensures that the gas moves slowly all the way down to
the BH horizon.  In the Igumenshchev et al. (2003) simulation, not
only is the radial velocity small ($\epsilon\sim0.1$), the tangential
velocity is also very small since the gas loses its angular momentum
efficiently through drag on the poloidal magnetic field.  Thus, the
gas has very little kinetic energy, and essentially all its rest mass
energy appears in other forms, e.g., heat or radiation or a
mechanical/magnetic outflow.  This leads to a high efficiency, just as
in the Geroch-Bekenstein engine.  The only question is what fraction
of the released energy actually escapes to infinity.

Inside $r_m$, where the gas breaks up into magnetically confined blobs
(or streams as in the Igumenshchev et al. 2003 MHD simulations), the
energy released will go partly into heating the blobs and partly into
heating the surrounding medium, which we call a ``corona.''  Let us
assume that the blobs are optically thick and that nearly all of their
heat energy is emitted as radiation.  To avoid complications, assume
that the radiation comes out isotropically in the rest frame relative
to the BH (we ignore Doppler beaming since the velocity is small).  In
addition, some of the energy that is deposited in the corona may also
be radiated locally and isotropically.  We refer to all of this
emission as ``disk radiation.''  Let us say that a fraction $f_{\rm
disk}$ of the energy released at each radius is emitted as disk
radiation.  For simplicity, we assume that $f_{\rm disk}$ is
independent of $r$.

For a Schwarzschild BH, a fraction $(1-\cos\delta)/2$ of isotropically
emitted radiation at radius $r$ escapes to infinity if $r<3/2$ (radii
inside the circular photon orbit) and a fraction $(1+\cos\delta)/2$
escapes if $r>3/2$, where $\cos\delta = [ 1 - (27/4r^2)
(1-1/r)]^{1/2}$.  Integrating over the differential energy release
$dE(r)/dr$ (eq.[1]) as a function of radius, we find that a fraction
$0.4375 f_{\rm disk}$ of the rest mass energy of the accreting gas
escapes to infinity as radiation, and a fraction $0.5625 f_{\rm disk}$
of the rest mass energy falls into the BH.  The precise value $0.4375$
depends on the particular assumptions we have made, but the general
result, that nearly half the radiated energy escapes to infinity, is
probably quite general.  Note that, as in eq. (1), all our
calculations are done in terms of the energy as measured at infinity.
Therefore, gravitational redshift effects are automatically included
in the estimates.

The remainder of the energy $(1-f_{\rm disk})$ goes into the corona
and comes out partly as a thermal wind and partly as Poynting flux.
Because both of these fluxes flow along the magnetic field lines, all
of this energy escapes to infinity.  Some of the energy may actually
come out as radiation, e.g., beamed from the outflowing wind or
created farther out where the wind meets an external medium.  Let a
fraction $f_{\rm jet}$ of the rest mass energy of the accreting gas
come out in this form of ``jet radiation,'' all of which escapes to
infinity.  Finally, a fraction $(1-f_{\rm disk}-f_{\rm jet})$ remains
as mechanical or magnetic energy and flows out into the external
medium.

Energy is, of course, conserved in this entire process.  If we start
with a parcel of gas of mass $m$ around a BH of mass $M$ and if the
gas accretes via a MAD flow as described above, then, after the gas
has fallen into the BH, the mass of the hole will increase to
$M+0.5625f_{\rm disk}m$, and an energy $(1-0.5625f_{\rm disk})mc^2$
will return to infinity, of which $(0.4375f_{\rm disk}+f_{\rm
jet})mc^2$ will be in the form of radiation and the rest in a
non-radiative form.  We now define two efficiencies: $\eta_{\rm rad}$,
the ratio of radiative energy reaching infinity to the rest mass
energy of the inflowing gas, and $\eta_{\rm energy}$, the ratio of the
total energy reaching infinity (radiative, mechanical, magnetic) to
the rest mass energy.  From the previous discussion, we obtain
\begin{equation}
\begin{array}{l}
\eta_{\rm rad,MAD} = 0.4375 f_{\rm disk}+f_{\rm jet}, \\
\eta_{\rm energy,MAD} = 1 - 0.5625 f_{\rm disk}.
\end{array}
\end{equation}

BKR76 estimated $\eta_{\rm rad}\sim 50\%$ in their Newtonian model of
turbulent magnetized accretion, arbitrarily assuming that the
effective cut-off radius of the accretion flow is $1.5 R_S$.  Their
estimate is remarkablely close to the more quantitative result we
obtain here, which includes relativity and takes the cut-off radius to
be $1R_S$.

\section{Discussion and Conclusions}

The analogy between the Geroch-Bekenstein engine and the MAD model is
quite close.  In the former, a strong wire arrests the falling mass
$m$ and lowers it slowly into the BH, while at the same time
transporting energy out to infinity.  The MAD model does the same
thing, except that the poloidal magnetic field plays the role of the
wire.  The model is thus a practical realization of the
Geroch-Bekenstein engine.  A problem with Bekenstein's (1972) original
proposal is that no physical wire is strong enough to survive the
intense tidal force of a BH (Gibbons 1972).  However, a poloidal
magnetic field with a topology similar to that in the MAD model can
circumvent Gibbons' argument (Znajek 1976).

Soltan (1982) has shown that, by comparing the integrated luminosity
of QSOs with the BH mass density in the local universe, one can
estimate the average efficiency of accretion in QSOs.  Elvis, Risaliti
\& Zamorani (2002) and Yu \& Tremaine (2002) have used this method to
argue for highly efficient accretion, with $\eta_{\rm rad}>0.1$ (see
also Merritt \& Ferrarese 2001).  In view of the conservative
assumptions made by Yu \& Tremaine (2002), and especially the results
shown in their Fig. 4, it would appear that the efficiency may be
closer to 1 than 0.1.  The relevant efficiency when considering the
Soltan argument is not $\eta$ as defined in the present paper, but the
ratio $\eta{'}$ of the radiative energy reaching infinity to the mass
energy added to the BH.  For the MAD model described here,
\begin{equation}
\begin{array}{l}
\eta{'}_{\rm rad,MAD} = \left(0.4375 f_{\rm disk}+f_{\rm jet}\right)/
0.5625 f_{\rm disk} = \\
\qquad \qquad \qquad 0.7778 + 1.7778 \left(f_{\rm jet}/ f_{\rm disk}\right).
\end{array}
\end{equation}
The model appears to have a large enough radiative efficiency to
satisfy the QSO observations.

We have dealt exclusively with the extraction of rest mass energy from
the accreting gas; to emphasize this point, we have focused on a
non-rotating BH in our discussion.  With a rotating BH, one can in
addition extract energy from the BH (e.g., Blandford \& Znajek 1977;
Li 2000; Punsly 2001; and references therein).  The presence of this
additional source of energy is a complicating factor when comparing
our model to others in the literature.  In the following, we focus
only on those published models in which the BH spin parameter $a/M$
remains the same before and after the accretion, so that we can be
certain that none of the released energy comes from the BH spin.

A standard thin accretion disk around a Schwarzschild BH ($a/M=0$) has
$\eta_{\rm rad} = 0.057$, while Thorne's (1974) limiting Kerr BH with
$a/M\approx 0.998$ has $\eta_{\rm rad} \approx 0.30$.  Li \&
Paczy\'nski (2000) have described a model in which a BH alternates
between episodes of accretion spin-up and spin-down, giving a net
efficiency of $\eta_{\rm rad} = 0.436$.  Gammie (1999) has discussed a
model in which magnetic fields exert a torque on an accretion disk at
the ISCO, and shows that $\eta_{\rm rad}$ is nearly 0.2 for a
Schwarzschild BH under ideal conditions and about $0.3-0.4$ for an
equilibrium Kerr BH with $a/M \approx 0.7$.  Agol \& Krolik (2000)
include the effects of returning radiation and show that the maximum
efficiency for a BH in spin-equilibrium is 0.36.  The MAD model
compares favorably with all these models (see eqs. [3], [4]).
Interestingly, the large efficiency of the MAD model is obtained with
a generic magnetic field configuration and for a non-rotating BH.

The MAD model is based on a key assumption, namely that radiatively
efficient disks drag in external magnetic field to the center.  We
await further theoretical and numerical work on this conjecture.  The
model also assumes that the gas inside the magnetospheric radius moves
slowly relative to free-fall.  Although supported by recent 3D MHD
simulations (Igumenshchev et al. 2003) and other discussions in the
literature (e.g., Ikhsanov 2001), this assumption needs to be tested
further.


We thank Jacob Bekenstein, Gena Bisnovatyi-Kogan, Bohdan
Paczy{\'n}ski, Brian Punsly and the referee for useful comments.  RN
was supported in part by NASA grant NAG5-10780.  IVI was supported by
the U.S. Department of Energy (DOE) Office of Inertial Confinement
Fusion under Cooperative Agreement No. DE-FC03-92SF19460, the
University of Rochester, the New York State Energy Research and
Development Authority.  MAA was supported by the Swedish Research
Council grant 650-19981237/2000.



\end{document}